\documentclass[showpacs,preprintnumbers,amsmath,amssymb,prc,floatfix,reprint]{revtex4-1}

\usepackage{graphicx}
\usepackage{color}
\usepackage{dcolumn}
\usepackage{bm}
\usepackage{longtable}
\usepackage{epsfig}
\usepackage{times}

\begin{document}

\title{Coherent control of the cooperative branching ratio for nuclear x-ray pumping}

\author{Adriana \surname{P\'alffy}}
\email{Palffy@mpi-hd.mpg.de}

\author{Christoph~H. \surname{Keitel}}
\email{Keitel@mpi-hd.mpg.de}

\author{J\"org \surname{Evers}}
\email{Joerg.Evers@mpi-hd.mpg.de}

\affiliation{Max-Planck-Institut f\"ur Kernphysik, Saupfercheckweg~1, 
69117 Heidelberg, Germany}

\date{\today}

\begin{abstract}
Coherent control of nuclear pumping in a three level system driven by x-ray light  is investigated. In single nuclei, the pumping performance is determined by the branching ratio of the excited state populated by the x-ray pulse. Our results are based on the observation that in ensembles of nuclei, cooperative excitation and decay leads to a greatly modified nuclear dynamics, which we characterize by a time-dependent cooperative branching ratio.
We discuss prospects of steering the x-ray pumping by coherently controlling the cooperative decay. First, we study an ideal case with purely superradiant decay and perfect control of the cooperative emission. A numerical analysis of x-ray pumping in nuclear forward scattering with coherent control of the cooperative decay via externally applied magnetic fields is presented. Next, we provide an extended survey of  nuclei suitable for our scheme, and propose proof-of-principle implementations already possible with typical M\"ossbauer nuclear systems such as $^{57}\mathrm{Fe}$.  Finally, we discuss the application of such control techniques to the population or depletion of long-lived nuclear states.
\end{abstract}





\maketitle


Coherent control in quantum optics and atomic physics provides an efficient tool to investigate atomic properties and favorably manipulate the dynamics of the system alike. Similar possibilities with nuclear systems have been considered with great interest \cite{Rivlin, ValiVali,Baldwin1,Baldwin2,Marcuse,nqo1,nqo2,wente} already shortly after the realization of the first laser in the optical band \cite{Maiman}.
However, many quantum optical control schemes require the effective coupling of the driving field to the considered transition (Rabi frequency) to be of the same order as the relaxation rate of the same transition \cite{Ficek,Scully}. Consequently it is very demanding to exploit the direct laser driving of nuclear systems experimentally. Furthermore, the dream of the nuclear laser is at present equally out of reach. The pursuit of coherent sources for wavelengths around or below 1~nm is supported however by the advent and commissioning of x-ray free electron lasers \cite{slac,xfel}, the
availability of which will stimulate the transfer of quantum optical schemes to nuclei along these lines.  

A different route to coherent control  of nuclear dynamics which does not rely on forthcoming x-ray light sources is coherent light scattering off nuclei in the low-excitation limit~\cite{Tramell_book,AK,Bible,Ralf,Bernhard}. In particular, coherent nuclear forward scattering (NFS) is a routine technique experimentally studied in many labs around the world. In NFS, high-frequency light such as that from a synchrotron radiation (SR) source is monochromatized at a nuclear resonance energy, and then scatters coherently off a nuclear target. As has recently been realized, while being conceptionally different from the attempts to directly transfer quantum optical schemes to the nuclear realm, NFS does allow to explore coherent control of nuclei in experimental settings already available today~\cite{Shvydko_MS,ScienceRalf,PRL103,JMOproc}.
The possibility of control exploited in these works arises from the fact that the resonant scattering off the nuclear ensemble (for instance identical nuclei in a crystal lattice) occurs via an  excitonic state, i.e., an excitation coherently spread out over a large number of nuclei.  In case of coherent scattering, the nuclei return to the initial state after scattering, such that the scattering path is unknown. This leads to cooperative emission, with scattering only in forward direction (except for the case of Bragg scattering~\cite{Kagan,Bible,Smirnov}) and decay rates modified by the formation of sub- and superradiant states as key signatures. 
\begin{figure}[b]
\begin{center}
\includegraphics[width=0.9\columnwidth]{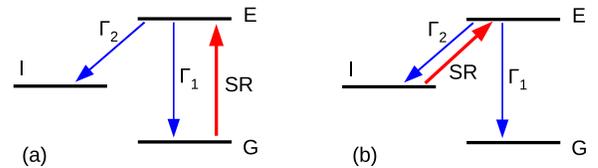}
\caption{\label{Lambda_scheme} The considered three level system. (a) State  $|E\rangle$ is populated by the SR pulse and can decay to the ground state  $|G\rangle$ or to the isomeric state  $|I\rangle$, presumed to be long-lived. (b) The SR pulse couples states  $|I\rangle$ and  $|E\rangle$. The initial state  $|I\rangle$ is assumed to be metastable. For both (a) and (b), the natural decay rates for the $|E\rangle \rightarrow |G\rangle$ and $|E\rangle \rightarrow |I\rangle$ transitions are $\Gamma_1$ and $\Gamma_2$, respectively. $|I\rangle$ and $|G\rangle$ could also be realized as different hyperfine sublevels of a single nuclear state, as it is often the case in atomic quantum optics. This allows for proof-of-principle implementations in  $^{57}\mathrm{Fe}$. }
\label{lambda}
\end{center}
\end{figure} 
Since due to the narrow linewidth of the nuclear transitions the life time of the exciton is sufficiently long, the latter can  be externally manipulated before it decays. This has been exploited in the experiment reported in~\cite{Shvydko_MS}, in which a suitable rotation of the direction of an applied magnetic field was used to suppress the coherent decay over a certain period. The observed storage of energy in an excitonic state was achieved by inducing destructive interference among the coherent decay channels~\cite{Shvydko_MS,Shvydko_theory}, in direct analogy to electromagnetically induced transparency~\cite{RevModPhys.77.633} and spontaneous emission suppression via spontaneously generated coherences~\cite{Ficek} well known from atomic quantum optics. Next to this externally induced interference, also the cooperative nature of NFS was exploited in a recent experiment~\cite{ScienceRalf}, which could experimentally verify the presence of a cooperative Lamb shift theoretically predicted in~\cite{PRLScully} for the case of light scattering off of atoms. These findings invite the search for further applications of coherent control of cooperative nuclear decay, and it has already been suggested by the present authors that this technique can be used to generate single photon entanglement in the x-ray regime~\cite{PRL103,JMOproc}.  

Motivated by this, in the present paper we investigate the coherent control of nuclear x-ray pumping in NFS. Nuclear x-ray pumping  denotes here the controlled transfer of population between different nuclear states by application of x-ray light. As model system, we assume nuclei in a three-level $\Lambda$-configuration as shown in Fig.~\ref{lambda}(a). The nuclei initially reside in the ground state $|G\rangle$. A SR pulse in NFS setup excites part of the nuclei to the excited state $|E\rangle$, followed by decay  either back to the initial state $|G\rangle$ or to the target state $|I\rangle$, which could be an isomeric state. In an isolated nucleus, the final state populations are governed by the branching ratio. We show that in ensembles of nuclei, cooperative light emission in NFS leads to an effective branching ratio. This {\it cooperative branching ratio} can significantly deviate from the single-nucleus branching ratio, and is time-dependent. By coherently suppressing  the cooperative decay,  the cooperative branching ratio can be controlled. 

We first analyze x-ray pumping in the ideal case of perfectly superradiant cooperative decay that can be completely suppressed at will after the excitation occurred. For this case, we find that our approach allows to increase the target state population by a so-called superradiant enhancement factor which is sample-dependent and potentially large. Then, we turn to a numerical analysis of NFS with realistic parameters, and discuss the performance and limitations in this particular implementation of x-ray pumping and cooperative branching ratio control. We find that in this setup, even though cooperative branching ratio  control and consequently target state population enhancement can be observed, the achievable enhancement factor  is only on the order of 2. Our analysis shows that the difference to the ideal case mainly arises from the dynamical population of sub-radiant states in the cooperative nuclear decay. 

We round up our analysis by an extended survey of suitable nuclei to implement our scheme. First, we propose two proof-of-principle experiments which could allow to observe the branching ratio control already using the traditional 14.4 keV 
M\"ossbauer transition in  $^{57}\mathrm{Fe}$. We then discuss the known M\"ossbauer nuclei for possible implementations with nuclear level systems involving a metastable state.

Efficient nuclear state population control is an essential ingredient in advanced measurement schemes such as state-selective nuclear scattering, as well as in applications like nuclear lasers or the manipulation of isomers. In particular, isomeric triggering or depletion---release on demand of the energy stored in the excited metastable nuclear state---has been proposed via a number of nuclear excitation mechanisms, such as photoabsorption \cite{Kara_Carr1,Carr_Kara}, Coulomb excitation \cite{Hayes,Kara_Carr5} or coupling to the atomic shells \cite{isomersPRL,Tkalya,palffyCP}. The triggering methods make use of a nuclear three-level scheme similar to  the one presented in Fig.~\ref{lambda}(b). The isomer essentially does not decay directly, but rather has to be excited to a trigger state $|E\rangle$ which can subsequently decay with a branching ratio also to the nuclear state $|G\rangle$--- here either the true ground state or a state directly connected to it. Control of the collective branching ratio in the case of NFS provides therefore a way to enhance the isomer depletion or population. Our results apply equally well to both scenarios in Fig.~\ref{lambda}.

The paper is organized as follows. In Section \ref{bc-ideal} we introduce the cooperative branching ratio and discuss an ideal case of branching ratio control. In Section~\ref{implementation}, we apply these results to an implementation based on coherent control of NFS, and present a numerical treatment of the coherently scattered light for realistic parameters. Section~\ref{BR} presents results for the cooperative branching ratio and nuclear state population dynamics taking into account the coherent decay and its possible suppression for the NFS case. Finally, Sec.~\ref{Conclusions} discusses and summarizes the results. Atomic units ($\hbar=1$) are used throughout the paper.

\section{\label{bc-ideal}Branching ratio control}

\subsection{The cooperative branching ratio \label{coll_BR}}
An isolated nucleus in the excited state  $|E\rangle$ can decay via radiative decay 
or  by internal conversion decay (IC) to either of the ground and intermediate state $|G\rangle$ and $|I\rangle$. The total  width of the excited state is then determined by the sum of the individual transition rates 
\begin{align}
\Gamma_{E} & =\Gamma_1+\Gamma_2 \nonumber\\
&= \Gamma_1^{\gamma}+\Gamma_1^{IC}+\Gamma_2^{\gamma}+\Gamma_2^{IC}\, ,
\end{align}
where the superscripts $\gamma$ and IC refer to the radiative and IC decay channels, respectively.
An important parameter is the branching ratio of a single nucleus
\begin{align}
\label{b-single}
b_1=\frac{\Gamma_2^{\gamma}+\Gamma_2^{IC}}{\Gamma_{E}}\,,
\end{align}
which gives the fraction of the excited nuclei that will decay to the $|I\rangle$ state. $\Gamma_E=\Gamma_1+\Gamma_2$ is the  natural decay rate of the excited state $|E\rangle$, i.e, the total incoherent decay rate.

For a collection of nuclei, the probability for radiative decay back to the initial state (the ground state $|G\rangle$)  can be greatly enhanced due to cooperative effects \cite{Bible}. For example, spatial coherence of the light source can lead to the formation of excitonic nuclear states, which are characterized by a delocalized excitation coherently shared by a large number of nuclei. The decay width of such excitonic states can be substantially larger than that of a single nucleus, leading to superradiant decay. Similarly, sub-radiant states with reduced decay rates may also exist. Cooperative decay occurs provided that the excitation cannot be localized at a single nucleus. This condition is met by coherent scattering, in which the initial and the final nuclear states coincide. In contrast, nuclear recoil, spin flip or the change of the nuclear state lead to localization of the excitation, and thus essentially to single-particle decay. In the case of the three level system sketched in Fig.~\ref{lambda}(a), coherent decay can occur only between states $|E\rangle$ and $|G\rangle$, provided that the initial and final magnetic sublevels coincide. The decay $|E\rangle\rightarrow|I\rangle$, on the other hand, as well as all IC channels and the radiative decay $|E\rangle\rightarrow|G\rangle$ involving spin flipping occur incoherently, with the natural decay rates. The cooperative decay can be characterized by an additional coherent contribution $\Gamma_c(t)$ to the decay rate from $|E\rangle$ to $|G\rangle$. Thus, when comparing the $|E\rangle \rightarrow |G\rangle$ and $|E\rangle \rightarrow |I\rangle$ transitions, the natural branching ratio $b_1$ no longer describes the fraction of excited nuclei that decay from $|E\rangle$ to $|G\rangle$ and to $|I\rangle$, respectively.

In order to account properly for the coherent  decay of the excitonic state $|E\rangle$, we introduce a  cooperative branching ratio. The starting point for defining this time-dependent quantity is the nuclear exciton decay, governed in the second Born approximation by the set of equations~\cite{Bible,Scully}
\begin{subequations}
\label{Bloch}
\begin{align}
\frac{d}{dt} P_e(t)
 &=  - [\Gamma_c(t)+\Gamma'_1+\Gamma_2]\,P_e(t)\, , \\
\frac{d}{dt}P_g(t) &= [\Gamma_c(t)+\Gamma'_1]P_e(t)\, ,  \\
\frac{d}{dt}P_i(t) &= \Gamma_2P_e(t) \, , 
\end{align}
\end{subequations}
where $P_e(t)$, $P_g(t)$ and $P_i(t)$ are the excited, ground and isomeric state populations at time $t$, respectively. The transition rate to the ground state contains the time-dependent coherent rate $\Gamma_c(t)$, while the decay rate of the excitonic state to the isomeric state $\Gamma_2$ is the same as for the case of a single nucleus. Note that the coherent decay rate $\Gamma_c(t)$ actually includes a part of the radiative width $\Gamma_1^{\gamma}$, namely the decays that proceed back to the initial magnetic sublevel. We therefore denote in the equation above by $\Gamma_1'$  the incoherent part of the radiative decay of $|E\rangle$ to $|G\rangle$, involving a nuclear spin flip, and the IC decay. 
The time-dependent transient cooperative branching ratio for the $|E\rangle\rightarrow|I\rangle$ transition is given by the ratio between the incoherent decay rate $\Gamma_2$ and the total decay rate for the excitonic state,
\begin{equation}
\label{b-c}
b_c(t)=\frac{\Gamma_2}{\Gamma_c(t)+\Gamma_1'+\Gamma_2}\, .
\end{equation}
A similar result has been obtained in the study of superfluorescent cascades in atomic excitations to Rydberg states \cite{Molander}.

\subsection{Control of the cooperative branching ratio\label{controlBR}}
Since immediately after the excitation, $\Gamma_c(t)$  is typically much larger than the incoherent decay rates $\Gamma_1$ and $\Gamma_2$, the cooperative branching ratio $b_c(t)$ during this time is considerably smaller than the single-particle branching ratio $b_1$ in Eq.~(\ref{b-single}). In effect, the time-integrated branching ratio in an ensemble of nuclei is {\it smaller} than the single particle branching ratio $b_1$, thereby reducing the efficiency of the optical pumping to the target state $|I\rangle$. In the following, we will study prospects for controlling the cooperative branching ratio $b_c(t)$ assuming that the cooperative light emission rate $\Gamma_c(t)$ can be manipulated. This can be achieved, e.g., by using the magnetic switching technique demonstrated in Ref.~\cite{Shvydko_MS}, or by externally destroying the spatial coherence throughout the lifetime of the excitonic state. In Sec.~\ref{implementation}, we will discuss the implementation of the suppression of $\Gamma_c(t)$ using these methods and analyze the performance of the branching ratio control for realistic parameters.

In the second Born approximation used in Eq.~(\ref{Bloch}), the time dependence of the coherently scattered light intensity can be related to the coherent width by \cite{Bible} 
\begin{eqnarray}
I(t)&=&\Gamma_c(t) P_e(t)
\nonumber \\
&=&N_0\Gamma_c(t)\mathrm{e}^{-\tilde{\Gamma}(t)t}\, .
\label{exciton}
\end{eqnarray}
Here, the number of nuclei excited by the SR pulse is denoted by $N_0$, and  we have introduced the effective decay rate \cite{Bible} 
\begin{equation}
\label{g-tilde}
\tilde{\Gamma}(t)=\frac{1}{t}\int_0^{t} \left(\Gamma_c(s)+\Gamma_1'+\Gamma_2\right)\:ds\, .
\end{equation}
Imagine now that we can switch off the coherent decay $\Gamma_c(t)$ beginning with time $t_s$. From Eq.~(\ref{g-tilde}), the total decay rate of the excited state can then be written as 
\begin{equation}
\label{eff-gam}
\tilde{\Gamma}(t)=\frac{1}{t}\int_0^{t} \left[ \Gamma'_1+\Gamma_2+\Gamma_c(s)\Theta(t_s-s)\right]\:ds\, .
\end{equation}
The population of the two final state levels $|G\rangle$ and $|I\rangle$ and the cooperative branching ratio can be obtained then by solving numerically the  equations (\ref{Bloch}) with the controlled coherent decay rate $\Gamma^s_c(t)=\Gamma_c(t)\Theta(t_s-t)$. Note that because the suppressed coherent decay includes a subchannel of the incoherent radiative decay, the suppression may influence the branching ratio of the transitions even at times for which the magnitude of the coherent decay diminishes. This is however only relevant when the radiative decay channel is not dominated by the IC channel. 

To explore the potential of the branching ratio control, we now assume ideal conditions of immediate suppression of the coherent decay and constant superradiant decay rate,
\begin{subequations}
\label{ideal_par}
\begin{align}
t_s &= 0\,,\\
\Gamma_c(t) &= \xi (\Gamma'_1+\Gamma_2) \approx \xi \Gamma_E\,,
\end{align}
\end{subequations}
with $\xi \gg 1$. The quantity $\xi$, identified as the dimensionless effective thickness parameter in Section \ref{time_dep}, is defined as 
\begin{align}
\label{def-xi}
\xi= \frac 14 \sigma_R N L\,,
\end{align}
where $\sigma_R$ is the radiative nuclear resonance cross-section, $N$ the number density of M\"ossbauer nuclei in the sample $L$ the sample thickness $L$.
Using the expressions (\ref{ideal_par}), the cooperative branching ratio without control of the superradiant decay evaluates to
\begin{align}
b_c^{NC} = \frac{\Gamma_2}{(\xi+1)(\Gamma'_1+\Gamma_2)}\,,
\end{align}
whereas the one with control of the superradiant decay (NSR) is
\begin{align}
b_c^{C} = \frac{\Gamma_2}{\Gamma'_1+\Gamma_2}\,.
\end{align}
Thus, the suppression leads to an increase of the population in the target state by a factor
\begin{align}
b_c^{C}/b_c^{NC} = \xi +1\,.
\end{align}
In conclusion, collectivity leads to the enhancement of the total target state population---in our case the isomeric state $|I\rangle$---in two ways. First, it accounts for an enhanced upper state population immediately after the SR pulse that will eventually decay to the $|G\rangle$ and $|I\rangle$ states. Second, as it follows from the equation above, the switching leads to a further enhancement by the  factor $\xi+1$, where $\xi$ is  proportional to the density of nuclei in the sample.

\section{\label{implementation}Implementation of branching ratio control in nuclear forward scattering}
In this section we  discuss a concrete model system for the  branching ratio control  implementation  in nuclear x-ray pumping presented in the previous section. For this, we consider the scattering of synchrotron radiation on M\"ossbauer nuclei embedded in a crystal target. For M\"ossbauer nuclei, the coherent nuclear excitation occurs without a localized recoil, and both the  duration and the transit time of the SR pulse shining on a  crystal target are short compared to the excited state lifetime $\tau$  of the nuclear excited state. The pulse therefore creates a collective nuclear excited state  which is a spatially coherent superposition of the various excited state hyperfine levels of a large number of nuclei in a certain coherence volume in the crystal. This coherence leads to acceleration of both excitation and deexcitation of the nuclei, as required for the branching ratio control. 
For definiteness, we focus in the theoretical description on the $\Lambda$-configuration depicted in Fig.~\ref{lambda}(a).

\subsection{Wave equation for the coherently scattered light}
The coherently scattered synchrotron light can be described with a wave equation similar to the atomic case \cite{Shvydko_theory}
\begin{equation}
\left(\nabla -\frac{1}{c^2}\frac{\partial^2}{\partial t^2}\right)\vec{E}=\frac{4\pi}{c}\frac{\partial}{\partial t}\vec{I}\, ,
\end{equation}
where $\vec{E}$ is the electric field component of the light, and $\vec{I}$ the nuclear source current. In slowly varying envelope approximation, and for light propagating in $y$ direction, this simplifies to an equation for the envelopes $\vec{\mathcal{E}}$ and $\vec{\mathcal{I}}$  given by 
\begin{equation}
\frac{\partial}{\partial y}\vec{\mathcal{E}}=-\frac{2\pi}{c}\vec{\mathcal{I}}\, .
\end{equation}
Calculating the nuclear source current in second order in the interaction of the light and the nuclei, one finally obtains a wave equation
\begin{equation}
\frac{\partial\vec{E}(y,t)}{\partial y}=-\sum_{\ell} K_{\ell} \vec{J}_{\ell}(t)\int_{-\infty}^t d\tau \vec{J}^{\, \dagger}_{\ell}(\tau)\cdot \vec{E}(y,\tau)\, .
\label{wave_eq}
\end{equation}
Here, the excitation and decay steps of the resonant scattering are represented by the nuclear transition current matrix elements $\vec{J}_{\ell}(t)$, and $\ell$ is a summation index running over all possible transitions, with properties characterized by $K_{\ell}$, and the nuclear sites \cite{Shvydko_theory}.  We assume for simplicity in our calculation only one nuclear site (for a more quantitative calculation taking into account particular nuclear sites or sample characteristics, see, e.g., the MOTIF code \cite{Motif}).   The equation above can be iteratively solved starting from an initial synchrotron radiation pulse $\vec{E}(t)=\vec{\mathcal{E}}_0 \delta (t)$ which is instantaneous on the time scale of the nuclear dynamics. The result is a sum
\begin{equation}
\vec{E}(y,t)=\sum_{n=0}^{\infty}\vec{E}_n(y,t)\, ,
\label{E_sum}
\end{equation}
in which each term represents a multiple scattering order. The total intensity of the scattered radiation due to the coherent decay channel is then given by  $I(t)=|\vec{E}(L,t)|^2$, where $L$ is the thickness of the sample. Since the incident SR term $\vec{E}_0$ only plays a role at $t=0$, we neglect it in the calculation of the intensity.

\subsection{Time-dependent coherent decay rate \label{time_dep}}
Let us first consider the case when the nuclear levels have no hyperfine splitting, and there is only one transition driven by the SR pulse, corresponding to a single $\ell$ in the sum in Eq.~(\ref{wave_eq}). For such a simple system  the expression for the intensity of the coherently scattered light can be obtained analytically \cite{Shvydko_theory,Bible},
\begin{equation}
I(\tau)=\mathcal{E}_0^2\xi \frac{\mathrm{e}^{-\tau}}{\tau} \left [J_1(\sqrt{4\xi \tau}) \right]^2\, .
\label{intensity}
\end{equation}
Apart from the dimensionless effective thickness $\xi$ defined in Eq.~(\ref{def-xi}), we have  also introduced  a dimensionless time coordinate $\tau=\Gamma_Et$. The Bessel function of first order $J_1$ can be approximated for the limit of very small times $\tau\lesssim 1/ \xi$  as
\begin{equation}
\frac{[J_1(\sqrt{4\xi \tau})]^2}{\xi\tau}\simeq \mathrm{e}^{-\xi\tau}\, ,
\end{equation}
such that immediately after the excitation, the decay of the excitonic state is exponential and faster than the natural decay,
\begin{equation}
I(\tau)=\mathcal{E}_0^2\xi^2 \mathrm{e}^{-(\xi+1)\tau}\, .
\label{superradiant}
\end{equation}
This shows that indeed the coherent decay is accelerated relative to the incoherent, spontaneous decay $\mathrm{e}^{-\tau}$ immediately after the excitation. The decay is superradiant, as assumed for the ideal case in Eqs.~(\ref{ideal_par}). We can now identify the enhancement factor $\xi$ introduced in Section \ref{controlBR} and previously denoted as superradiant in the introduction as the effective thickness of the sample,  proportional to the sample nuclei density. 

\begin{figure}[t]
\begin{center}
\includegraphics[width=0.9\columnwidth]{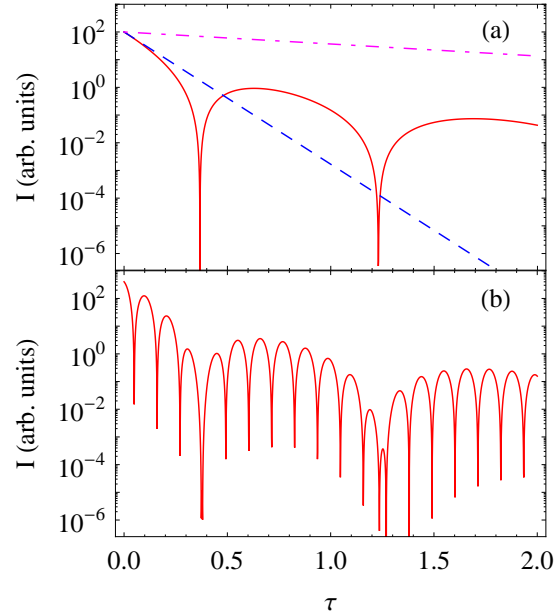}
\caption{\label{intensities}(a) Intensity of the coherent scattered light (solid red line),  incoherent natural decay (dash-dotted magenta line), and superradiant decay in Eq. (\ref{superradiant}) (dashed blue line) for  a nuclear sample of effective thickness $\xi=10$ and a single nuclear transition driven by the SR pulse. (b) The hyperfine splitting of the nuclear levels produces quantum beats in the intensity of the scattered light. The quantum beat frequency is given by the hyperfine energy correction $\Omega_0=28.3\Gamma$, where $\Gamma$ is the natural width of the excited state. } 
\end{center}
\end{figure}

For delayed times, $\tau\gg 1/ \xi$, the decay has a completely different character. In contrast to the initial superradiant decay, the fall-off is slow and presents the onset of dynamical beats, determined by the Bessel function. A typical time response for $\xi=10$, together with the incoherent natural decay rate is shown in Fig.~\ref{intensities}(a). The dynamical beat arises in our approach of the scattering problem from the multiple scattering terms in Eq.~(\ref{E_sum}). In terms of the exciton created by the SR pulse, the dynamical beat  can be explained as interference effects of the radiative eigenmodes of the crystal. The exciton itself can be written as a Bloch wave, which is generally not a radiative normal mode of the crystal, but rather a superposition of the eigenmodes that have a spread in eigenfrequencies and decay rates. Since the eigenmodes are not Hermitian orthogonal, interference effects lead to the appearance of the dynamical beats in the evolution of the exciton  \cite{Bible}.

The coherent width of the excitonic state $\Gamma_c$ is related to the effective thickness parameter $\xi$.  For times immediately after the SR pulse, the width of the state is constant and given by $\Gamma_c=\xi\Gamma_E$. However, for later times $\Gamma_c$ becomes time-dependent, and its value can  be calculated numerically from the time-dependent intensity of the scattered light. For this, we evaluate from Eq.~(\ref{exciton})
\begin{align}
\frac{\dot{I}(t)}{I(t)}  = \frac{\dot{\Gamma}_c(t)}{\Gamma_c(t)}- [\Gamma'_1 + \Gamma_2 + \Gamma_c(t)]\,,
\end{align}
where the dot denotes differentiation with respect to time, leading to
\begin{align}
\dot{\Gamma}_c(t) = \left[ \frac{\dot{I}(t)}{I(t)} + \Gamma'_1 + \Gamma_2 + \Gamma_c(t)\right]\,\Gamma_c(t)\,. \label{g-num}
\end{align}
This allows to obtain $\Gamma_c(t)$ from the numerically calculated intensity $I(t)$.

If  hyperfine splitting of the nuclear levels or shifts between the levels of nuclei located in different chemical sites are present, light of different frequency components is scattered. This results in a quantum beat modulation of the cooperative decay rate, varying periodically between superradiant and subradiant emission into the coherent channels, with beat frequencies determined by the frequency differences among the various transition resonances. In Fig~\ref{intensities}(b) we show as an example the intensity of the scattered light as a function of time for a nuclear system with total angular momenta of the ground and excited states 1/2 and 3/2, respectively.  In this case the SR pulse drives the $M1$ transition between the hyperfine sublevels $\Delta m=0$. The quantum beat frequencies were chosen  $\Omega_0=\pm 28.3\Gamma_1$, corresponding to realistic parameters for the M\"ossbauer transition from the ground state to the first excited state in $^{57}\mathrm{Fe}$.

It should be mentioned that  the expression in Eq.~(\ref{intensity}), first derived to describe the propagation and multiple scattering of gamma rays in resonance with a nuclear absorption line, is applicable for excitons in different physical systems and parameter regimes. For instance, it has also been derived to describe the propagation of short weak laser pulses through resonant matter \cite{Crisp} and has been used experimentally to enhance transient absorption in the infrared \cite{Silberberg}. Experimentally, the coherent propagation and quantum beats of quadrupole polaritons in $\mathrm{Cu_2 O}$ have been reported \cite{Frohlich}, and also the coherent control of excitons in semiconductor heterostructures has been intensely studied (see, e.g.,~\cite{PhysRevB.73.235345} and references therein).  The dynamical beat has been observed also in the coherent interaction of femtosecond extreme uv-light pulses with Helium atoms \cite{Thomas}. This has been interpreted as a light propagation effect arising from the dispersion  around an absorption resonance.  A light propagation formalism featuring the calculation of the refraction index in the nuclear medium has been applied also for NFS \cite{vanB92}, and is an equivalent approach to the one presented in this work.


\subsection{Control of the coherent decay via magnetic switching}
The key to control the cooperative branching ratio is the coherent decay of the excitonic state. Control of the coherent decay of the nuclear exciton has been already used to  produce  gamma echos in NFS experiments originally using M\"ossbauer sources \cite{gammaecho}. Subsequently  nuclear exciton echos produced by ultrasound vibrations of the sample in NFS of SR  were observed \cite{ultrasound}. The coherent decay was  manipulated in this experiment via the relative phase between the electromagnetic field scattered from two sample foils. Alternatively, changing the hyperfine magnetic field at the nuclear sites also provides a way to control the coherent decay. Following the experiment described in Ref.~\cite{JPhysCondM} on the effect of an abrupt reversal of the hyperfine magnetic field direction for NFS of light from a M\"ossbauer source, results confirming the feasibility of nuclear coherent control also in NFS of SR were presented in Ref. \cite{Shvydko_MS}. The decay rate of $^{57}\mathrm{Fe}$ nuclei in a $^{57}\mathrm{FeBO}_3$ crystal excited by 14.4~keV SR pulses was changed by switching the direction of the crystal magnetization. The nuclear hyperfine fields were used to partially switch the coherent decay channel of the nuclear excitation off and subsequently on, demonstrating the possibility to store nuclear excitation energy.

In the following we consider the possibility to suppress the coherent decay for our branching ratio control scheme via the magnetic switching technique described in Ref.~\cite{Shvydko_MS}. A switching of the magnetic field direction changes the quantization axis for the nuclear system. If such a rotation is applied almost instantaneously and directly after the SR excitation, the quantization axis changes and the population of the hyperfine levels is redistributed according to the new hyperfine basis. For specific switching times and rotation angles, the transition amplitudes for the coherent decay can destructively interfere, thus suppressing the coherent decay. The partial suppression and subsequent release of the coherent nuclear decay are the consequence of interference between the hyperfine transitions, bearing a close resemblance to the underlying effect of electromagnetically induced transparency in quantum optics \cite{Scully,RevModPhys.77.633}.

A detailed analysis of the effect of the  switching time on the coherent decay intensity and polarization for the $M1$ M\"ossbauer transition in iron $^{57}\mathrm{Fe}$ has been carried out in Ref.~\cite{JMOproc}, following the original idea in Ref.~\cite{Shvydko_MS}. It has been shown that an almost complete suppression of the coherent decay can be achieved for certain geometry  configurations, if the magnetic field direction is rotated from  parallel to the sample to being perpendicular to the sample and parallel to the light propagation direction at the time moment when the quantum beat is at its minimum~\cite{Shvydko_MS,JMOproc}. Assuming that originally only $\Delta m=0$ transitions were excited, the complete suppression of the first order coherent scattering can be obtained by rotating the magnetic field at the switching time $t_s=(n-1/2)\pi/\Omega_0$, where $n\in\{1,2,\ldots\}$ and $\Omega_0$ is the hyperfine energy correction for the originally driven $\Delta m=0$ transitions. 

Experimentally, the magnetic switching is facilitated in crystals that allow for fast rotations of the strong crystal magnetization via weak external magnetic fields. For iron, one of the most suitable host materials is $\mathrm{FeBO}_3$,  a canted antiferromagnet with a plane of easy magnetization parallel to the (111) surfaces. Initially, a constant weak magnetic field induces a magnetization parallel to the crystal plane surface and aligns the magnetic hyperfine field $\vec{B}$ at the nuclei. The hyperfine field is strong, on the order of 30~T, leading to a pronounced Zeeman shift of the magnetic sublevels.  The magnetic switching is then achieved by a  pulsed magnetic field in a perpendicular crystal plane that rotates the magnetization by an angle $\beta$ and realigns the hyperfine magnetic field.  Because of the perfection of the crystal, the desired rotation of the magnetization occurs abruptly, over less than 5~ns~\cite{Shvydko_EPL}.
The effective decay rate in Eq.~(\ref{eff-gam}) then becomes 
\begin{equation}
\tilde{\Gamma}(t)\simeq \frac{1}{t}\int_0^{t} \left(\Gamma'_1+\Gamma_2+\Gamma_c(s)\Theta(t_s-s)\right)\:ds\, .
\end{equation}
In the equation above, we use "$\simeq$" instead of "$=$" because the switching is not perfect and only the leading first order scattering of the coherent decay is suppressed. 

Compared to the ideal conditions of Eqs.~(\ref{ideal_par}),  the realistic case presents  two important differences. First, as it has been discussed in Section \ref{time_dep}, the coherent decay is no longer superradiant for $\tau\gg 1/\xi$. The magnetic switching is therefore most efficient when performed as soon as possible after the excitation, during the time span of the superradiant coherent decay. Second, it is not possible to suppress the coherent decay immediately after the excitation, and thus $t_s>0$. For the already addressed case of the M\"ossbauer transition in $^{57}\mathrm{Fe}$, the first moment when the magnetic switching is possible with complete suppression of the coherent decay is given by the first quantum beat minimum $t_s=\pi/(2\Omega_0)\simeq 8$~ns. The same scheme can be used for any $M1$ transition, independent of the ground and excited states spins, by using an appropriate geometry that allows the SR pulse to drive only the $\Delta m=0$ transitions. To some degree, the shortest possible switching time can be adjusted by controlling the magnitude of the hyperfine magnetic field at the nuclei, for instance by cooling the sample. Nevertheless, these two differences will lead to reduced enhancement of the branching ratio  compared to the ideal case.

\subsection{Survey of suitable M\"ossbauer nuclei}

At present there are more than 40 isotopes with transitions for which the M\"ossbauer effect has been observed, featuring a number of different nuclear level configurations. However, most of the research on this topic is related to the $^{57}\mathrm{Fe}$ isotope and its traditional M\"ossbauer 14.4~keV transition, which does not offer a low-lying metastable state.
In the following we outline possible implementations of our setup in two different cases, depending on the available nuclear levels.

\subsubsection{Nuclear level configurations based on hyperfine states}

The first range of possible implementations only require a ground and an excited nuclear level, each split up into hyperfine states. This requirement is already met by the M\"ossbauer first excited state of $^{57}\mathrm{Fe}$. One type of proof-of-principle experiment is based on the observation that the cooperative branching ratio control discussed here relies on the suppression of coherent decay, leading  to an effective enhancement of the incoherent decay. For example, in Fig.~\ref{Lambda_scheme}(a), the coherent decay $|E\rangle \to |G\rangle$ is suppressed such that more nuclei decay incoherently to the target state $|I\rangle$. It is therefore  possible to demonstrate  branching ratio control not only by probing the change in the nuclear state population, but also by measuring the incoherent decay. The enhancement of incoherent decay associated with the branching ratio control is accessible even for simplified level schemes using the M\"ossbauer 14.4 keV transition in $^{57}\mathrm{Fe}$. For this, one could  measure the rate of photons emitted in a direction perpendicular to the incident SR beam, since the incoherent photons are emitted in all spatial directions, while coherently scattered photons  in forward direction only. Alternatively, one could measure the incoherent decay by monitoring the electrons emitted in the non-radiative process of IC. Such measurements are possible, e.g., via a multi-channel plate \cite{SturhahnPRB}. In either case, a relative enhancement of the number of incoherent photons or IC electrons should be observed when the coherent decay is switched off, thereby verifying the concept of branching ratio control. 

A more direct alternative implementation  in $^{57}\mathrm{Fe}$ relies on monochromatized  light sources that render hyperfine state-selective excitations  possible (see for instance Ref.~\cite{SmirnovPRB}). Such a pump beam could be used to selectively excite the transition $m_g=1/2 \to m_e=1/2$ in $^{57}\mathrm{Fe}$.
While the coherent decay from the excited $m_e=1/2$ level occurs to $m_g=1/2$ only, the incoherent decay leads to both $m_g=\pm 1/2$. We can therefore identify $m_g=1/2$ with $|G\rangle$, $m_e=1/2$ with $|E\rangle$, and $m_g=-1/2$ with $|I\rangle$,  obtaining thus a configuration as in Fig.~\ref{Lambda_scheme}(a).
The finite lifetime of the metastable state $|I\rangle$ in this scenario can be identified with thermal spin flips between the two ground hyperfine states in $^{57}\mathrm{Fe}$, which should be kept low by cooling the sample. The branching ratio control would effectively lead to a larger pump rate into the ground state $m_g=-1/2$. The ratio between the two ground state populations could be probed by applying a broadband pulse exciting both the $m_g=1/2 \to m_e=1/2$ and the $m_g=-1/2 \to m_e=-1/2$ transitions. The different transition frequencies lead to beats in the NFS time signal. Since the population ratio in the two ground states $m_g=\pm 1/2$ directly affects the amplitudes of the two interfering scattering channels, the population difference in the two hyperfine ground states can be observed as a change in the temporal beats of the NFS time signal. This measurement technique could be further improved by applying a microwave pump field between the two ground states to empty one of the two ground states before initiating the nuclear x-ray pumping~\cite{PhysRevLett.69.2815,Vagizov}. This way, one could measure the number of pumped nuclei relative to zero, rather than relative to the total number of nuclei as it is the case for an initially balanced population distribution.

\subsubsection{Nuclear level configurations based on isomeric states}

Apart of iron, other nuclides suitable for M\"ossbauer spectroscopy should possess excited nuclear states with lifetimes in the range of  $\mu$s to approx. 10~ps, and transition energies  between  5 and 180 keV. Longer (shorter) lifetimes than indicated lead, according to the Heisenberg uncertainty principle, to too narrow (broad) emission and absorption lines, which no longer effectively overlap. Transition energies beyond  180 keV cause too large recoil effects which destroy the resonance, while gamma quanta with energies smaller than  5 keV are mostly absorbed in the source and absorber material. In order to find out whether a particular nuclear transition proceeds recoilless, one should calculate the Lamb-M\"ossbauer factor, which can be approximated in the Debye model as \cite{Barb}
\begin{equation}
f_{LM}=\textrm{exp}\left\{-\frac{2E_R}{k_B\theta_D}\left(1+4\frac{T^2}{\theta_D^2}\int_0^{\frac{\theta_D}{T}} \frac{xdx}{e^x-1}\right)\right\}\, .
\label{LambMoss}
\end{equation}
Here, $k_B$ is the Boltzmann constant, $\theta_D$ the Debye temperature, $E_R$ the recoil energy of the transition and $T$ the temperature of the sample. The Lamb-M\"ossbauer factor determines the probability that the recoilless absorption and emission occurs without exciting lattice vibrations and changing the state of the particular nucleus. The closer to one $f_{LM}$ is, the larger the fraction of recoilless excitation in NFS. Enriched iron  $^{57}\mathrm{Fe}$ at room temperature, for instance, has $f_{LM}=0.804$ \cite{SturhahnPRL}.

Among the M\"ossbauer elements that have been experimentally confirmed,  eight of them have more than one M\"ossbauer transition and present an interesting three-level system for the cooperative branching ratio control  as the ones shown in Fig.~\ref{Lambda_scheme}. This would allow for instance for enhanced storage of energy in an isomeric state, or isomer depletion. However, none of the levels involved are really long-lived and therefore do not present the incentive of isomeric state population. It is possible that other nuclear transitions of the M\"ossbauer isotopes with energies and lifetimes within the required parameters might be also proceeding recoillessly. In particular interesting are the  M\"ossbauer nuclei which present an isomeric state, such as $^{189}\mathrm{Os}$ or $^{178}\mathrm{Hf}$. The $^{178\rm{m2}}{\rm Hf}$ isomer is a high-energy long lived isomer with $\tau=31$~y and $E=2.4$~MeV \cite{toi}. The isomer's conveniently long lifetime and high excitation energy of 2.4~MeV make it particularly attractive for the study of possible energy release on demand. 
Until now, the attempts to trigger the energy release from the 31-year  $^{178\rm{m2}}{\rm Hf}$  isomer via broadband SR  have been a  highly controversial issue \cite{Collins,Ahmad1,Ahmad2,Carroll}.

The $^{189\rm{m}}\mathrm{Os}$ 30.814~keV isomer with natural lifetime $\tau_0=5.8$~h has possible triggering levels at 97.35~keV and 216.663~keV. Out of these, the transition to the 97.35~keV is more likely to be recoilless due to its smaller  energy. Osmium's Debye temperature is somewhat uncertain, with tabulated values of $\theta_D=411\pm 94$~K \cite{OsDebye}. Assuming a value of 500 K, we obtain using Eq.~(\ref{LambMoss}) a Lamb-M\"ossbauer factor of only 0.1. In the case of $^{178\rm{m2}}{\rm Hf}$, the isomer has an experimentally confirmed triggering level at 2573.5 keV, 126.1 keV above the metastable state. The excitation of the isomer to this level would however not proceed recoillessly due to the high transition energy.
A controversial low-lying triggering level at about 40~keV above the isomer observed by Ref.~\cite{Collins} in triggering experiments using broadband SR light could not be confirmed by other groups \cite{Ahmad1,Ahmad2}.  If such a level exists, precise knowledge of its position is desirable for the efficient coherent excitation of the triggering transition via monochromatized SR light.  For the case of confirmed triggering transitions, the Lamb-M\"ossbauer factor for  energies of around 40 keV reaches for hafnium a value of approx. 0.2. Since both osmium and hafnium have rather small 
Lamb-M\"ossbauer factors, one can envisage implanting the isomers into a host material with higher Debye temperature. Estimating the recoilless fraction of absorption and emission in nuclear transitions for impurities in hard crystalline host materials requires however dedicated calculations.

Another  practical issue is whether a fast efficient magnetic switching is possible in crystals containing  $^{178\rm{m2}}{\rm Hf}$  or $^{189\rm{m}}\mathrm{Os}$ isomers. Although the advantageous $\mathrm{FeBO}_3$ crystal can only provide fast magnetic switching for $^{57}\mathrm{Fe}$, other magnetic materials may be used for different nuclei as $^{189\rm{m}}\mathrm{Os}$ or $^{178\rm{m2}}{\rm Hf}$. Thin films in multilayer structures with high coercitivity, for instance, allow for a good control of the crystal magnetization. 
Due to specific layer couplings, multilayer and superlattice systems can exhibit a richness of magnetic properties that is not found in bulk materials \cite{Ralf}.  Very thin layers of almost all transitive metals can be deposited on superpolished wafers by rf-magnetron sputtering in a rare gas atmosphere \cite{Ralf,RalfPRL89}.  
 Depending on what nucleus is envisaged, the host material with proper magnetic properties should be sought for. Also, depending on the multipolarity and the nuclear state spins of the involved levels, the geometry and the switching parameters have to be investigated, following the procedure described in Ref.~\cite{JMOproc}.

\begin{figure}[t]
\begin{center}
\includegraphics[width=0.9\columnwidth]{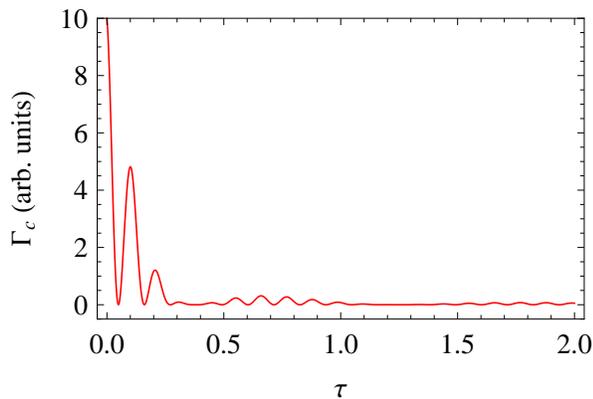}
\caption{\label{int/gamma} The  time-dependent coherent width $\Gamma_c(\tau)$ corresponding to the scattered  intensity spectrum presented in Fig.~\ref{intensities}(b). See text for further explanations.  } 
\end{center}
\end{figure} 
%

\section{\label{BR}Results}

To assess the performance of cooperative branching ratio control and state-selective x-ray pumping in NFS, we have investigated a general test case of a nuclear three level system with a $M1$ $|G\rangle\leftrightarrow|E\rangle$ transition driven by the SR pulse, as depicted in Fig.~\ref{lambda}(a). The third intermediate level is assumed to be metastable. We are interested to find out which fraction of the originally excited nuclei  $|E\rangle$ reach the isomeric state $|I\rangle$ and how does their number depend on the sample properties. The nuclear level population dependence  on the dimensionless time $\tau$ can be calculated from Eqs. (\ref{Bloch}). For this general case we have made the approximation $\Gamma_1'=\Gamma_1$. 
Assuming an initial geometry of the setup such that the SR pulse only drives the $\Delta m=0$ transitions, there are mainly two nuclear parameters that determine the scattered light intensity and subsequently the pumping performance. One of them is the hyperfine energy correction $\Omega_0$, and the other one is the natural, incoherent branching ratio $b_1$, see Eq.~(\ref{b-single}). Additionally, the effective thickness of the sample, corresponding to the number of M\"ossbauer nuclei present in the sample, also plays an important role, as it will be discussed later on. Although the effective thickness depends on the radiative nuclear resonance cross section $\sigma_R$, $\xi$ is rather considered to be a sample than a nuclear parameter.

In Fig.~\ref{int/gamma} we present  the  time dependence of the  coherent decay  width $\Gamma_c(t)$ as obtained numerically according to Eq.~(\ref{g-num}) from the intensity of the scattered radiation in  Fig.~\ref{intensities}(b). While the coherent width is large close to $\tau=0$, it becomes negligible afterwards. The quantum beat due to the presence of two driven hyperfine transitions  in the intensity appears also in the time dependence of the coherent decay width.
 
\begin{figure}[t]
\begin{center}
\includegraphics[width=0.9\columnwidth]{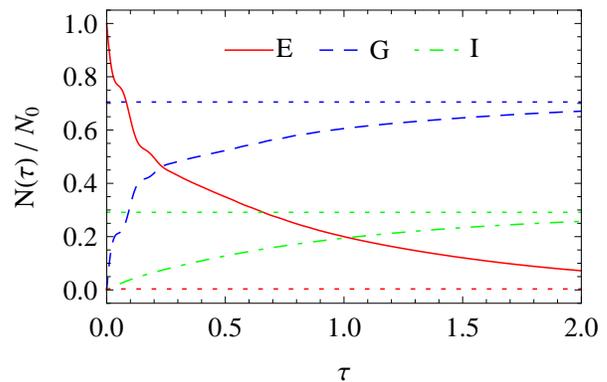}
\caption{\label{no_switching} The population $N(\tau)/N_0$ of the three nuclear levels as a function of the dimensionless time parameter $\tau$ for unperturbed coherent decay. The single-nucleus branching ratio was taken for this case $b=0.5$ and the effective thickness of the sample $\xi=10$. The horizontal dotted lines indicate the steady state values. }
\end{center}
\end{figure}

The numerical results for the time dependence of the nuclear levels population are given in Fig.~\ref{no_switching} for the case of a natural, incoherent branching ratio of $b=0.5$, $\Omega_0=\pm 28.3\Gamma_E$ and $\xi=10$. 
The coherent decay only plays an important role immediately after the SR pulse, and $\Gamma_c(t)\simeq 0$ for increasing $\tau\gg 1/\xi$. For $\tau\lesssim 1/\xi$, practically all decay of the excited state $|E\rangle$ occurs to the ground state, with very small population of the isomeric state $|I\rangle$. However, at larger times the coherent decay is practically zero and the remaining excited nuclei decay to the $|G\rangle$ and $|I\rangle$ states according to the incoherent branching ratio.

\begin{figure}[t]
\begin{center}
\includegraphics[width=0.9\columnwidth]{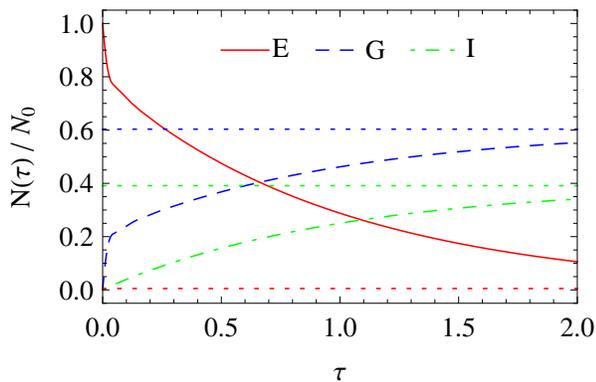}
\caption{\label{switching} The population $N(\tau)/N_0$ of the three nuclear levels as a function of time for suppressed coherent decay starting with $t_s=\pi/(2\Omega_0)$. The incoherent branching ratio is taken to be $b=0.5$ and the effective thickness of the sample is $\xi=10$. The horizontal dotted lines indicate the steady state values.}
\end{center}
\end{figure}

Magnetic switching  offers the possibility to reduce the effective period of the  coherent decay by suppressing it beginning with the first minimum of the quantum beat $t_s=\pi/(2\Omega_0)$. The time-dependent population of the three nuclear levels can be obtained from solving the set of equations (\ref{Bloch}) with the partly suppressed coherent width $\Gamma_c(t)\Theta(t_s-t)$. Numerical results are presented in Fig.~\ref{switching} for the same case of an incoherent branching ratio of $b=0.5$ and an effective thickness of the sample of $\xi=10$. We see that more nuclei decay to the isomeric state than in the case of unperturbed coherent decay presented in Fig.~\ref{no_switching}.

\begin{figure}[t]
\begin{center}
\includegraphics[width=0.9\columnwidth]{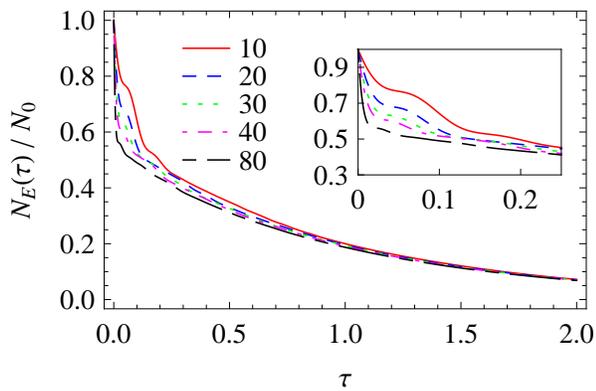}
\caption{\label{E_fct_x}  Excited state population for the unperturbed coherent decay (no switching) for several effective thickness parameters $\xi$.}
\end{center}
\end{figure} 

The effective thickness  $\xi$ (corresponding to the number of M\"ossbauer nuclei in the sample) 
 determines the coherent nuclear width $\Gamma_c(0)=\xi\Gamma_E$ at the moment of excitation by the SR pulse. Since the number of nuclei excited by the SR pulse depends on both the number of M\"ossbauer nuclei in the sample and on the width of the nuclear excited state, we have $N_0\sim \xi^2$. The coherent scattering off a thick sample is therefore more efficient in pumping the nuclear excited state $|E\rangle$. A larger effective thickness   increases  the number of excited nuclei in the sample and also the speed of the coherent decay. In order to separate these two aspects, we investigate the 
 dependence of the excited state population $N_E(\tau)/N_0$ on the thickness parameter $\xi$ in Fig.~\ref{E_fct_x} for the case of unperturbed coherent decay. The time-dependence of the excited state population for the thickness parameters $\xi=10,20,30, 40$ and $80$ is presented.  The main differences occur for small times $\tau$. At large times $\tau$,   only the incoherent decay determines the decay of the excited state, and the excited state population becomes almost the same for all considered $\xi$s. As it appears from 
Fig.~\ref{E_fct_x}, for the chosen effective thickness parameters, the coherent decay accounts only for the decay of about a half of the originally excited nuclei. This fraction is increasing with increasing $\xi$. The remaining  half of the excited state  population decays incoherently under the natural exponential decay law with the incoherent branching ratio $b$.

\begin{figure}[t]
\begin{center}
\includegraphics[width=0.9\columnwidth]{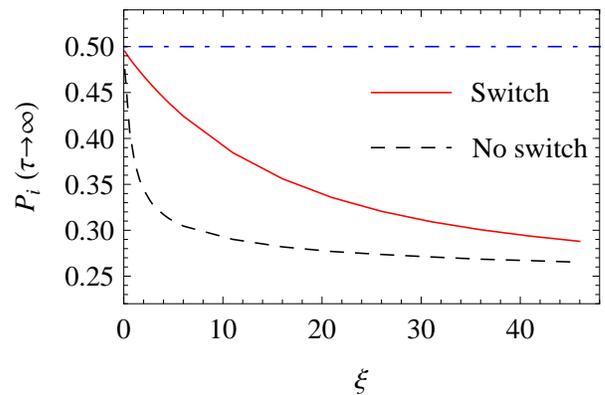}
\caption{\label{I_fct_x}  Steady-state isomeric state population plotted against the effective thickness parameters $\xi$. The black dashed line shows the case of unperturbed coherent decay without switching. The red solid line shows corresponding results with coherent control of the branching ratio. The blue dash-dotted line indicates the single-particle result without cooperative decay.}
\end{center}
\end{figure}

The population of the isomeric state $|I\rangle$ decreases with the effective thickness parameter, as shown in Fig.~\ref{I_fct_x}. This figure shows the steady state population of the isomeric state after the excited state population has completely decayed. However, the actual number of nuclei that have reached the isomeric state $|I\rangle$ is given by $N_0\,P_i(t)$ and will in fact increase with the thickness parameter $\xi$.  The linear dependence of $N_0$ on $\xi$  dominates the non-linear behavior shown in Fig.~\ref{I_fct_x}.

Let us now consider the case of magnetic switching suppression of the coherent decay. The new dependence of the isomeric state population $|I\rangle$ on $\xi$ is shown as solid red line in Fig.~\ref{I_fct_x}. Compared to the case of no switching, the isomeric state population for each $\xi$ is larger.  Nevertheless, with increase of the thickness parameter, the population of the isomeric state decreases as in the case without switching. Moreover, towards high $\xi$, the population of the isomeric state with and without switching becomes approximately the same. This is due to the relation between the sample effective thickness $\xi$ and the coherent decay speed-up. As discussed in Section~\ref{time_dep}, the coherent decay is superradiant only for times $\tau\lesssim 1/\xi$. The larger the thickness of the sample, the  faster is the  coherent decay extinguished, leaving active only the incoherent decay channels.  The effect on the switching at $\tau_s=t_s\Gamma_E=\pi/56.6=0.055$ 
is therefore less and less efficient for increasing $\xi$, since most of the superradiant decay occurs before the switching.

\begin{figure}
\begin{center}
\includegraphics[width=0.9\columnwidth]{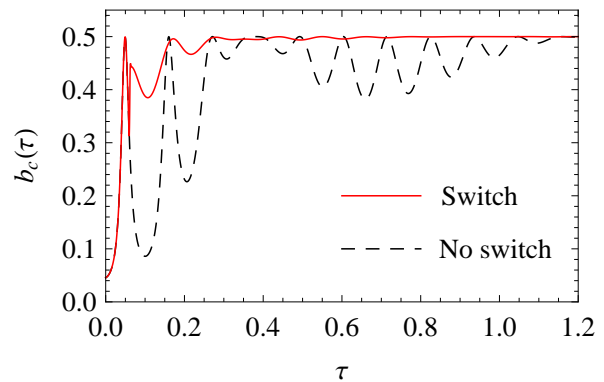}
\caption{\label{comparison} The cooperative branching ratio $b_c(\tau)$  corresponding to an incoherent branching ratio of $b=0.5$ with and without switching at $\tau=0.055$ for an effective thickness parameter of $\xi=10$. }
\end{center}
\end{figure}

Finally, in Fig.~\ref{comparison} we show a comparison of the cooperative branching ratio $b_c(\tau)$ calculated 
with and without switching at $t_s$  for $\xi=10$ and an incoherent branching ratio of $b=0.5$. Eventually, if by varying the hyperfine magnetic field (and consequently the hyperfine energy correction $\Omega_0$) the switching occurs immediately after the formation of the exciton and suppresses all the coherent decay, one obtains an enhancement factor of the isomeric state population of approx. 1.70 for $\xi=10$, 1.86 for $\xi=40$ and 1.93 for $\xi=80$. 

A more straightforward alternative for eliminating the coherent decay channel is to destroy the phase coherence of the exciton. While magnetic switching is only possible after a certain time $t_s$ such that a large fraction of the upper state population is always lost via cooperative decay, destroying the phase has the advantage that it can be achieved immediately after the excitation, minimizing thus the excitation loss.  To destroy the coherence,  a spatially inhomogeneous electric or magnetic field can be used, leading to a spatially inhomogeneous level shift and thus to a differential phase shift for the nuclei. In order to estimate the effect of an inhomogeneous field with correlation length $\rho$ on our nuclear sample, we assume that two nuclei with relative distance $d>\rho$ have random relative phase after interacting with this field, while two nuclei with relative distance $d\leq \rho$ are still in phase. We further assume that without the inhomogeneous field, a total coherence volume $V$ containing $N$ atoms leads to a cooperatively enhanced intensity $I_{coh}\sim N^2$. The coherence volume can be estimated from the spatial (transversal) coherence area of the incident SR light times a length $\lambda_\parallel$ mainly limited by absorption and scattering~(see Sec. 4.1.4 in Ref.~\cite{Bible}). In contrast, incoherent emission would scale linearly with $N$ as $I_{inc}\sim N$. The effect of the inhomogeneous field can then be modeled by splitting the volume $V$ into cubes with volume $\rho^3$. The nuclei within each cube are still in phase and thus emit cooperatively. In contrast, the nuclei of different cubes emit incoherently with respect to each other. Thus, after application of the inhomogeneous field, an intensity
\begin{align}
I_{\rho} &\sim N_c \cdot N_\rho^2 =  \frac{V}{\rho^3} \cdot \left (\frac{\rho^3}{V}\: N\right )^2 
= \frac{\rho^3}{V}\: N^2
\end{align}
with $N_c$ the number of cubes in $V$ and $N_\rho$ the number of atoms per cube with volume $\rho^3$ can be expected. The  application of the inhomogeneous field reduces then the cooperative emission as
\begin{align}
\label{scaling}
\frac{I_{\rho}}{I_{coh}} \sim \frac{\rho^3}{V}\,.
\end{align}
This estimate is valid as long as $\rho^3\leq V$, and  $\rho$ is sufficiently large such that many nuclei are contained within each cube of volume $\rho^3$. If the correlation length is small on the scale of the distance between adjacent nuclei, then the sample emits incoherently after the application of the inhomogeneous field, and $I_{\rho}/I_{coh}\sim 1/N$. As a rule of thumb, it is therefore sufficient to induce relative phase shifts covering the interval  $(0,2\pi)$ on a length scale determined by the coherence length inside the sample to considerably reduce the cooperative emission, and thereby affect the cooperative branching ratio. We confirmed the scaling in Eq.~(\ref{scaling}) by numerically simulating the effect of the inhomogeneous field on the cooperative emission for samples of up to few thousand nuclei.

Independently of the method used to eliminate the coherent decay, the obtained enhancement factor is in direct relation with the previously made observation  that the coherent decay accounts only for the decay of about a half of the originally excited nuclei. The main advantage for the pumping of the isomeric state $|I\rangle$ is therefore rather occurring due to the large coherent width $\Gamma_c(0)$ which  increases the excitation probability of the originally populated state $|E\rangle$. Since the effective thickness can take very large values (an effective thickness of $\xi=100$
corresponds to the actual sample thickness of about 20~$\mu$m in the case of iron $^{57}\rm{Fe}$), the enhancement factor can be substantial. However, the actual length $l_{c}$ in the sample thickness over which the coherent excitation can occur  is limited by scattering and absorption processes.  As an example,  the enhancement for the 14.4 keV resonance in $^{57}\mathrm{Fe}$ in NFS geometry is limited by photoabsorption to $10^3$ \cite{Bible}.

\section{Conclusions \label{Conclusions}}

In this paper we have studied the effects of  coherent resonant scattering of SR light off nuclei that present a three-level scheme with a metastable target state. The two investigated x-ray pumping configurations correspond to the population or the depletion of the long-lived nuclear state, respectively. The additional coherent decay channel that arises due to spatial coherence effects renders the  tabulated constant branching ratios that only include the incoherent decay channels obsolete.

We have introduced  a cooperative branching ratio for the three-level system, which monitors the evolution of the nuclear excited levels population and accounts for the additional time-dependence of the coherent decay of the exciton. This cooperative branching ratio is time-dependent and at short times after the excitation can be very different than the incoherent, natural branching ratio. The possibility to control this cooperative branching ratio and to increase the 
isomer population or depletion in our three level system by magnetic switching was investigated. 

In the ideal case, assuming purely superradiant decay of the nuclear exciton with constant rate and the possibility of instantaneous suppression of the coherent decay, the population in the target state can be enhanced via cooperative branching ratio control. The superradiant enhancement factor $\xi$ has been identified to be the effective thickness of the sample, limited only by the coherence length of the SR pulse.

Although at first sight promising, the enhancement brought by control of the coherent decay in the actual implementation in NFS of SR turns out to be only of a factor of approx. 2. The main reason for this is that the decay is only initially superradiant, and the suppression of the coherent decay is only possible after a minimum non-zero time. However, the creation of a nuclear exciton, which has as requirement the recoilless nuclear absorption and decay, can enhance the nuclear excitation probability by up to three orders of magnitude. The enhancement of the excitation probability itself is then reflected in the population of the other two nuclear states $|I\rangle$ and $|G\rangle$.
We conclude therefore that release on demand of the 
nuclear energy stored in  isomers is facilitated by coherence effects when occurring by driving a M\"ossbauer transition to a triggering level.  An experimental verification of the M\"ossbauer and magnetic switching properties of  
 nuclei in metastable states is the first step for  coherent control of nuclear state population and decay properties. In this direction, improvement in sample preparation and techniques related to thin films of radioactive atoms as host material open new possibilities in the investigation of exotic, unstable nuclei or isomeric states. In conjunction with the present overall trend to perform traditional nuclear and atomic physics experiments originally developed on stable nuclei  with radioactive, metastable or exotic nuclear species, such investigations are on their way.

\begin{acknowledgments}
We would like to thank Ralf R\"ohlsberger for helpful discussions.
\end{acknowledgments}

\bibliographystyle{apsrev}
\bibliography{palffy}

\end{document}